\documentclass[final,5p,times,twocolumn]{elsarticle}

 \usepackage{graphicx,psfrag}
\usepackage{amssymb}
\usepackage{lineno}

\journal{Nuclear Instruments and Methods in Research A}

\begin{document}

\begin{frontmatter}



\title{Dynamically polarized target for the $g_2^p$ and $G_E^p$ experiments at Jefferson Lab}


\author[JLab]{J. Pierce\corref{cor1}}
\ead{jpierce@jlab.org}
\author[UNH,MIT]{J. Maxwell}
\author[UNH]{T. Badman}
\author[JLab]{J. Brock}
\author[JLab]{C. Carlin}
\author[UVa]{D. G. Crabb}
\author[UVa]{D. Day}
\author[JLab]{C.D. Keith}
\author[UVa]{N. Kvaltine}
\author[JLab]{D.G. Meekins}
\author[UVa,UTK]{J. Mulholland}
\author[UVa]{J. Shields} 
\author[UNH]{K. Slifer}
\fntext[MIT]{Present address: Laboratory for Nuclear Science, Massachusetts Institute of Technology, Cambridge, MA 02139 USA}
\fntext[UTK]{Present address: Dept. of Physics and Astronomy, University of Tennessee, Knoxville, TN 37996 USA}
\cortext[cor1]{Corresponding author}

\address[JLab]{Thomas Jefferson National Accelerator Facility, Newport News, VA 23606, USA}
\address[UNH]{Dept. of Physics, University of New Hampshire, Durham, NH 03824, USA}
\address[UVa]{Dept. of Physics, University of Virginia, Charlottesville, VA 22904 , USA}

\begin{abstract}
We describe a dynamically polarized target that has been utilized for two electron scattering experiments
in Hall~A at Jefferson Lab.  The primary components of the target are a new, high cooling power $^4$He evaporation refrigerator, and 
a re-purposed, superconducting split-coil magnet.  It has been used to polarize protons in irradiated NH$_3$ at a temperature
of 1\,K and at fields of 2.5 and 5.0 Tesla.  The performance of the target material in the electron beam under these conditions will be discussed.  Maximum polarizations of 28\% and 95\% were obtained at those fields, respectively.  
To satisfy the requirements of both experiments, the magnet had to be routinely rotated between angles of 
0$^\circ$, 6$^\circ$, and 90$^\circ$ with respect to the incident electron beam.  
This was accomplished using a new rotating vacuum seal which permits rotations to be performed in only a few minutes.  
 \end{abstract}


\end{frontmatter}

%%\linenumbers

\section{Introduction} \label{Intro}
Dynamically polarized solid targets play an integral role in the physics program at Jefferson Lab.  To date, they have been
utilized on several occasions in experimental Halls B and C to examine the spin structure and electromagnetic
structure of both the proton and neutron, as well as the excited states of
the proton.  The targets operated in those halls have been described in separate 
articles~\cite{Averett99, Keith2003, Keith2012}, while the target described here marks the first use of a solid polarized target in
experimental Hall~A.  It mainly consists of components used previously in Halls B and C, heavily modified
to satisfy the requirements of the Hall~A experiments.  In addition, new components have been fabricated for improved
performance, reliability, and safety.

\section{Experimental Overview} \label{Overview}
\begin{figure}[t]
\begin{center}
\includegraphics[width=3in]{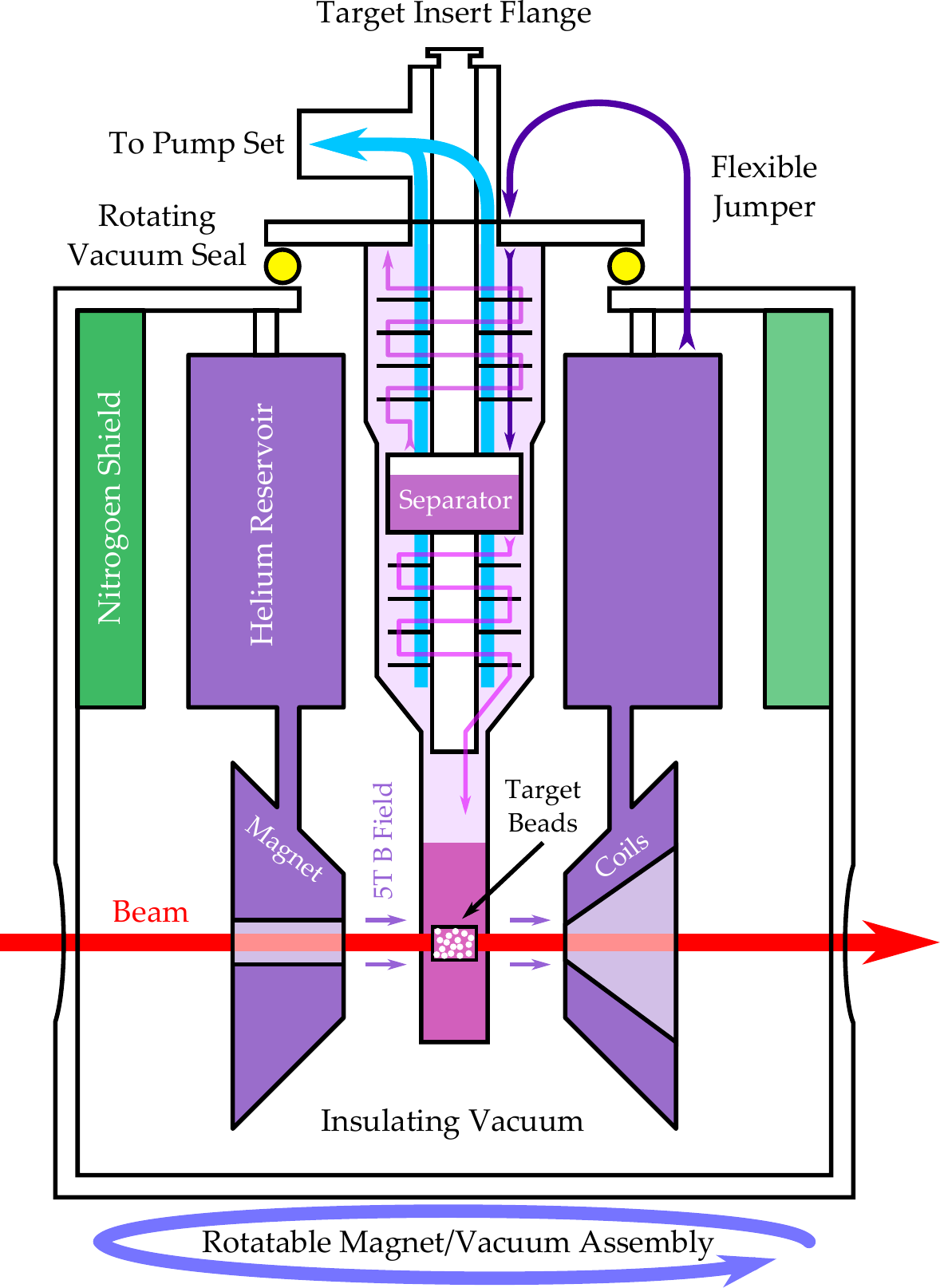}
\end{center}
\caption{Polarized target assembly for the $g_2^p$ and $G_E^p$ experiments.  The magnetic field is parallel to
the electron beam in this orientation.}
\label{Setup}
\end{figure}
Two separate experiments requiring a dynamically polarized proton target were approved for operation in Hall~A at Jefferson Lab.
The first of these experiments, referred to as ``$g_2^p$," aimed to measure the proton's transverse spin structure function $g_2^p$ at momentum-transfer squared
values as low as $Q^2 \lesssim 0.02$\,(GeV/c)$^2$ \cite{g2p}.  The second experiment, herein referred to as ``$G_E^p$,'' measured the
proton elastic form factor ratio $\mu G_E/G_M$ in the range $Q^2 = 0.01-0.7$\,(GeV/c)$^2$ \cite{GEp}.  
Both experiments examined the scattering of spin polarized electrons from spin polarized protons
at very forward angles. To extend the measurements to the lowest $Q^2$ values, a normally conducting 
septum magnet was located between the polarized target and the 
two Hall~A spectrometers to bend the most forward-going scattered electrons into the spectrometers.  
Both experiments proposed to use the polarized target system that had been utilized in Hall~C on
three previous occasions as well as at SLAC and is described by Averett {\em et al.} \cite{Averett99}.  This system features a high cooling power $^4$He evaporation refrigerator, a target insert accommodating multiple target samples, and 
a 5\,T superconducting split-coil magnet\footnote{Oxford Instruments plc} specifically designed for scattering experiments with a wide range of field directions.  

Because of their similar electron-beam energy requirements and because they shared much of the same equipment, 
the two experiments ran concurrently.  However, they required different 
values for directions of the proton polarization (and therefore the target's magnetic field).
For the $g_2^p$ experiment, this direction was 90$^\circ$ with respect to the incident electron beam, while 
for $G_E^p$ it was 6$^\circ$.  Additional measurements were made at 0$^\circ$ to measure $g_1^p$ for comparison to a result from Hall~B~\cite{Prok2009}.  
To accommodate alternation between the experiments at a given beam energy, field rotations were necessary as often as several times a week, so making each rotation as brief and reliable as possible was a priority.

In the Hall~C target system, rotation was accomplished by suspending the entire target cryostat from a large rotation stage.  
The 1\,K refrigerator and target insert also rotated along with the magnet, and the procedure required approximately eight hours
to perform successfully.  The vacuum pumps for the refrigerator had to be stopped and the refrigerator filled with a slight over-pressure
of helium gas before the plumbing between the pumps and refrigerator was disconnected.  After rotation, the
target insert was removed and placed back into the refrigerator, oriented along the beamline.  
Finally the plumbing was reassembled and the refrigerator cooled once more to 1\,K.

In Hall~A, the time required for this procedure has been dramatically reduced 
to only a few minutes by placing a new, rotary vacuum seal between the 1\,K refrigerator and the top of the cryostat, 
and by constructing a new plumbing manifold  between the refrigerator and the pumps.   
These allow the refrigerator and target insert to remain fixed with respect to the beam line
while the rest of the target assembly rotates. Since there is no need to remove the target insert or 
break the vacuum plumbing, the new scheme also reduces the risk of contaminating the helium system 
with air, or damaging the target samples during removal or insertion.

Finally, we note that the ammonia samples were polarized at both 5\,T and 2.5\,T fields.  The latter value was used during the $g_2^p$
runs at the lowest energies despite the lower polarizations obtained at this field. Here a transverse 5\,T field 
would have deflected scattered electrons outside the acceptance of the Hall~A spectrometers.  A 2.5 T field was chosen because of the availability of a 70 GhZ microwave circuit for the DNP process.  The deflection from this reduced field could be compensated for using chicane magnets upstream of the target.

\section{Polarized target system} \label{System}
Dynamic nuclear polarization (DNP) is a standard technique for producing polarized solid targets for nuclear and
particle experiments \cite{Crabb97}.  To realize DNP, a paramagnetic species in the form of
a free or unpaired electron is introduced into the target material, either by dissolving a
stable radical into the material (if the latter is liquid at room temperature), or by producing radicals 
directly within the material using ionizing radiation.  In this case, the target material consisted of irradiated NH$_{3}$.
The electrons are highly polarized by cooling the sample to a low temperature and exposing it to a high magnetic field.
For example, at the 1\,K and 5\,T operating conditions of this target, the electron polarization is 99.8\%.
Off-center microwave saturation of the radicals' Electron Spin Resonance (ESR) frequency is then used
to transfer their polarization to nearby nuclei, with one or more mechanisms, such as the solid effect, thermal mixing or the cross
effect, being responsible for the polarization transfer \cite{Maly2008}.
The polarization of nuclei near the paramagnetic radicals is transported throughout the bulk of the sample via spin diffusion,
and may be positive or negative, depending upon whether the microwave frequency 
is below or above the ESR frequency.    In well designed systems, proton polarizations exceeding 95\% \cite{Crabb1990} and
deuteron polarizations approaching 90\% \cite{Keith2012} have been achieved.

The general setup of the polarized target is shown in figure~\ref{Setup}.  The system consists of an aluminum scattering chamber
with multiple thin windows on its periphery for the entry and exit of the electron beam and for the exit of scattered particles.  
The scattering chamber is evacuated to approximately 10$^{-7}$\,torr  
and provides an insulating vacuum for the cryogenic components inside.  These consist of a liquid nitrogen heat shield, 
an 80 l liquid helium dewar, the superconducting magnet, and the 1\,K evaporation refrigerator.  The entire system hangs
from an aluminum and steel platform positioned at the pivot point between the two Hall~A spectrometers.  
A large rotary track allows the scattering chamber, LN$_2$ shield, 
LHe dewar, and magnet to rotate about their central axis.  The 1\,K refrigerator and target insert are rigidly attached to
the platform and interface with the top of the cryostat via a rotating vacuum seal.  They therefore remain fixed with respect
to the electron beam line as the other target components rotate underneath.

\subsection{Magnet} \label{Magnet}
\begin{figure}
\begin{center}
\includegraphics[width=2.75in]{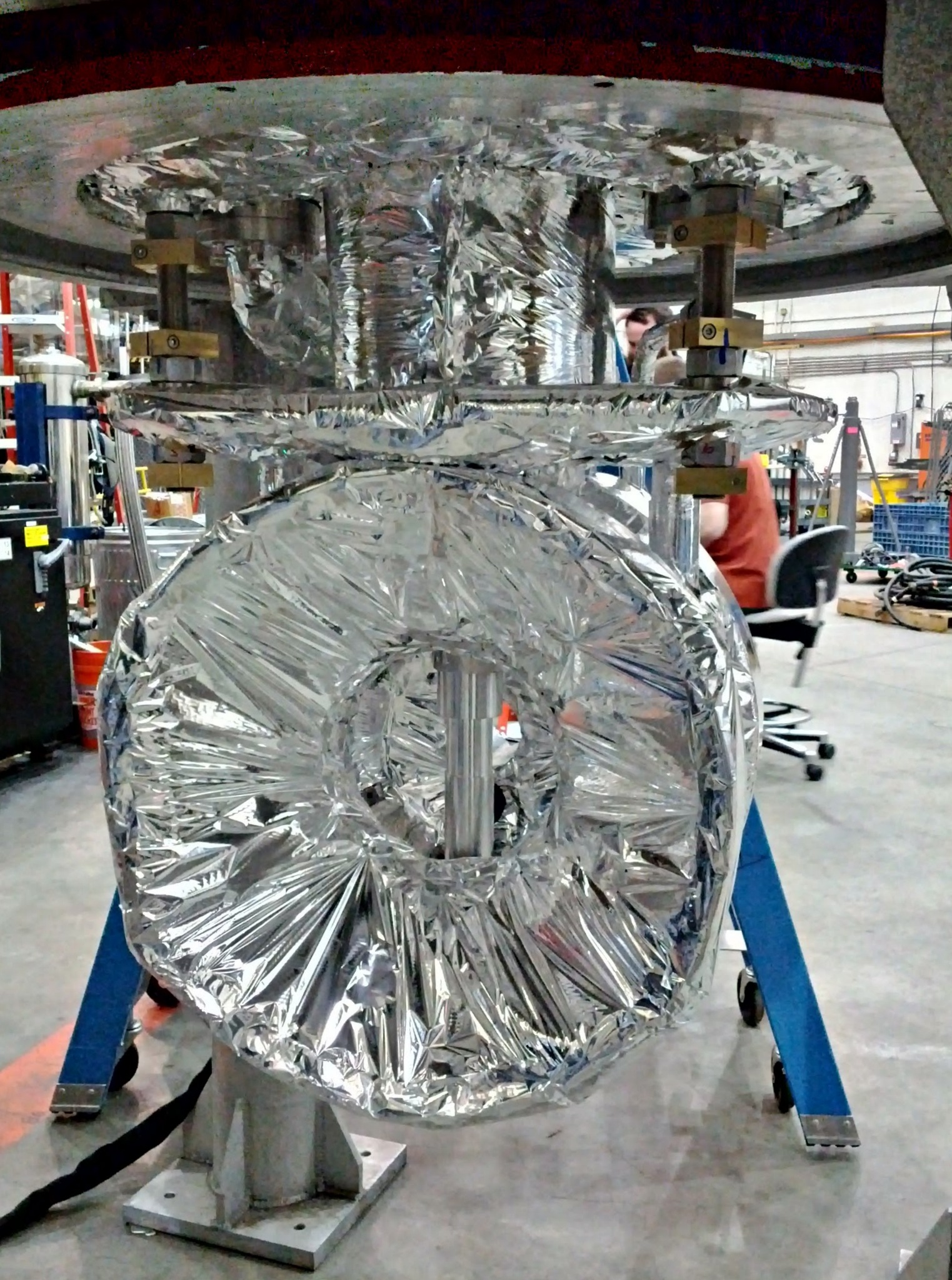}
\end{center}
\caption{Hall~B polarized target magnet suspended from the Hall~C polarized target cryostat, after covering with super-insulation.}
\label{Magnetfig}
\end{figure}
Failure of the Hall C magnet of reference~\cite{Averett99} necessitated a quick replacement with a 
set of 5\,T superconducting coils that were constructed\footnote{Oxford Instruments plc} for the Hall~B polarized target.
Figure~\ref{Magnetfig} shows a photograph of the Hall~B magnet suspended from the Hall~C helium dewar.
While the two magnets share some similarities, the Hall~B magnet was only intended to provide
a longitudinal polarizing field.  Therefore, differences do exist and are described below.

Both magnets have an 200\,mm open bore and an open angle of about $\pm17^\circ$ perpendicular to the field.
The Hall~C coils have a $\pm50^\circ$ open angle on both the up- and down-field sides of the magnet, whereas
the Hall~B coils have this open angle only on one side.  The quench protection circuitry for this
magnet is located on the opposite side, with a 75\,mm diameter port for the incident electron beam (in the Hall~C magnet, the quench protection is located above the coils in a separate stainless steel container).
The magnets have an 80\,mm diameter opening at the top for the target refrigerator and insert.
The Hall~C coils produce a field with a relative uniformity of $<10^{-4}$ over a spherical volume of 30\,mm diameter.
The uniform field region of the Hall~B magnet is smaller: $<10^{-4}$ over a cylindrical volume 20\,mm in diameter
and 20\,mm long.  While this is smaller than the dimensions of the ammonia target cells (25\,mm diameter $\times$
30\,mm long), we saw no adverse effects on the proton polarization.

Major modifications were necessary to install the Hall~B magnet in the Hall~C cryostat.  First, the magnet was rotated
180$^\circ$ about the field axis in order to locate the magnet leads at the top rather than the bottom of the
coil packages.  The access port for the leads was also used to supply liquid helium to the coils via a short length of stainless steel
hose connected to the underside of the cryostat's liquid helium dewar.  One aluminum support ring was attached to the
top of the magnet and a second suspended from the helium dewar using three 1-inch threaded rods.  With assistance
from Jefferson Lab's Survey and Alignment Group, the magnet was accurately positioned relative to the top plate of the
cryostat, and the two support rings were then clamped together.  We regard this as an improvement over the original
scheme, where the Hall~C magnet was rigidly suspended from the helium dewar with two indium-sealed flanges, and no fine positioning of the magnet inside the cryostat was possible.

\subsection{Refrigerator} \label{Refrigerator}
\begin{figure}[t]
\begin{center}
\includegraphics[width=2.1in]{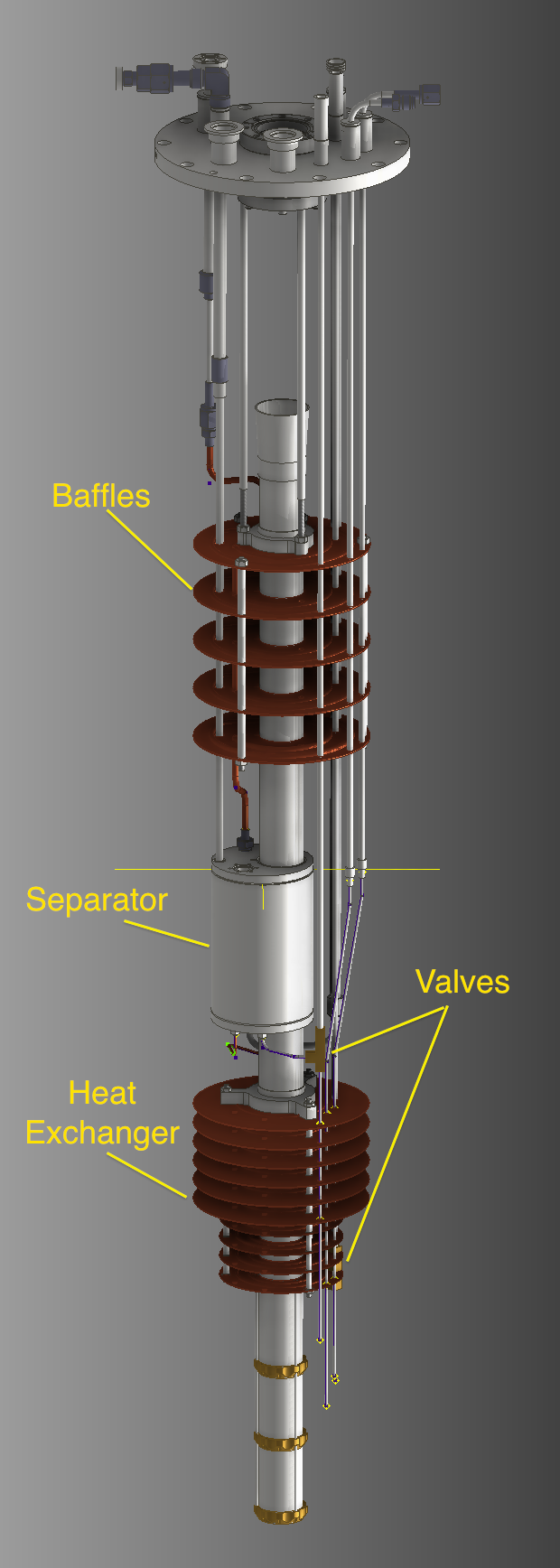}
\end{center}
\caption{The 1\,K evaporation refrigerator.}
\label{Fridge}
\end{figure}

A new, high cooling power, $^4$He evaporation refrigerator was constructed to replace the original refrigerator that
was damaged during its last use in Hall~C.
 The new refrigerator design includes modifications to
accommodate the new rotation scheme, improve reliability, and to satisfy the requirements of the ASME pressure vessel code.  The design, shown in figure~\ref{Fridge}, 
is well-established and will be briefly described here.

Using a well-insulated, flexible transfer line, 4\,K liquid helium is continuously siphoned from the
superconducting magnet's dewar into the top of a 1\,liter stainless steel vessel called the ``separator'',
where it drains through a 1\,mm thick plate of sintered stainless steel to remove
vapor that is transferred with the liquid.  The vapor is pumped away using a small 
diaphragm pump and cools a series of perforated copper baffles located
between the separator and the pumping manifold for the evaporation refrigerator.  A vapor
flow of 5\,slpm is sufficient to cool
the uppermost, warmest baffle to about 70\,K.  The separator is instrumented with a thermometer and a
miniature superconducting level probe.  

Liquid is drained from the bottom of the separator through a 3\,mm copper tube and delivered to a pumped bath of liquid helium
that is used to cool the polarized target samples to 1\,K.  Between the separator and 1\,K bath, 
the tube is thermally anchored to a second series of perforated copper plates which are cooled 
by gas pumped from the bath.  A small needle valve, located at the cold end of this gas-liquid heat exchanger,
is used to meter the flow of liquid to the bath. The valve is actuated by a room temperature stepper motor, and a
computer-controlled feedback loop is used to 
maintain a constant bath level without user intervention.
A second needle valve is used to bypass the heat exchanger for more rapid cooling.

All components for the refrigerator are constructed around a thin-walled, 2\,in stainless steel tube that helps guide
the target insert (see Section~\ref{Insert}) into the bath.  This tube extends several cm into the bath, and
is outfitted with a calibrated cernox resistor\footnote{Lake Shore Cryotronics, Inc.}, a $^3$He vapor pressure bulb, and a static tube for measuring
the $^4$He vapor pressure above the bath.  High-precision capacitance manometers are used for the 
vapor pressure measurements, which establish the sample temperature during the thermal
equilibrium calibration of the NMR system (see Section~\ref{NMR}).  A kapton film heater is attached near the bottom
of the guide tube and serves as an oven for annealing the beam-induced radiation damage 
to the ammonia samples (see Section~\ref{Anneals}).

\begin{figure}
\begin{center}
\includegraphics[width=3in]{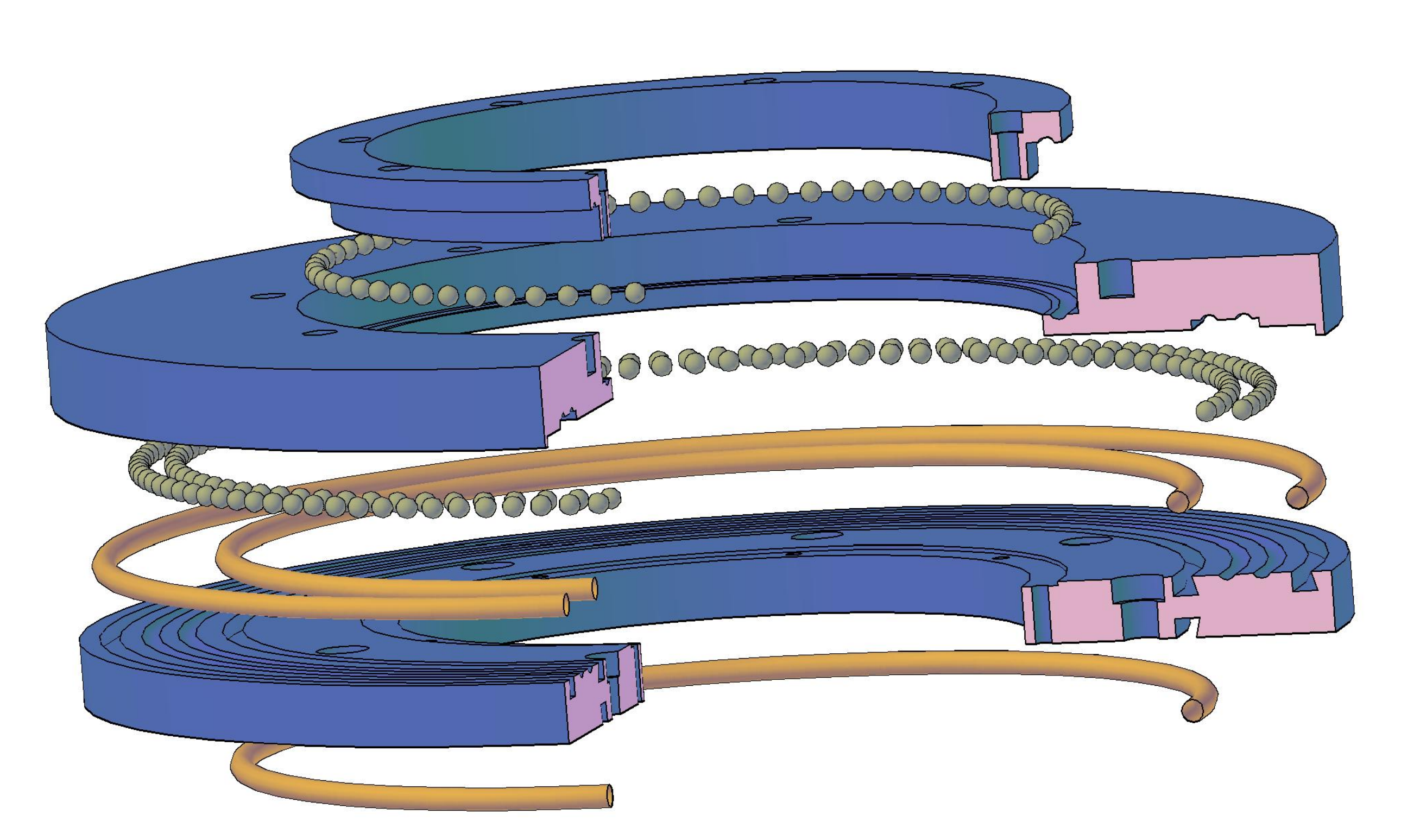}
\end{center}
\caption{Rotating vacuum seal for the refrigerator pumping tube.  The bottom flange seals against the
top plate of the target cryostat using a preexisting bolt circle and O-ring.  The center flange is free to rotate against this
flange with two O-rings and steel bearing races.  Differential pumping between the two O-rings with a small vacuum pump
is necessary to ensure vacuum integrity during rotation only.}
\label{Seal}
\end{figure}

The refrigerator slides into a 7\,in diameter stainless steel pumping tube that is suspended from the
top plate of the target cryostat.  A 46\,cm long, 60\,mm diameter aluminum extension attaches to the bottom of
the pumping tube using standard knife-edged flanges.  The central part of this extension is machined to
a wall thickness of 0.10\,mm, for beam entrance and exit.  A custom-designed rotating vacuum seal is
located between the upper end of the pumping tube and the cryostat's top plate.  This seal is shown in figure~\ref{Seal}.
Using a 12,000\,m$^3$hr$^{-1}$ Roots pump set, the refrigerator has a base temperature of about 0.9\,K, and a cooling power
of approximately 3\,W at 1.4\,K.

\subsection{Target insert} \label{Insert}

\begin{figure}
\begin{center}
\includegraphics[width=3.4in]{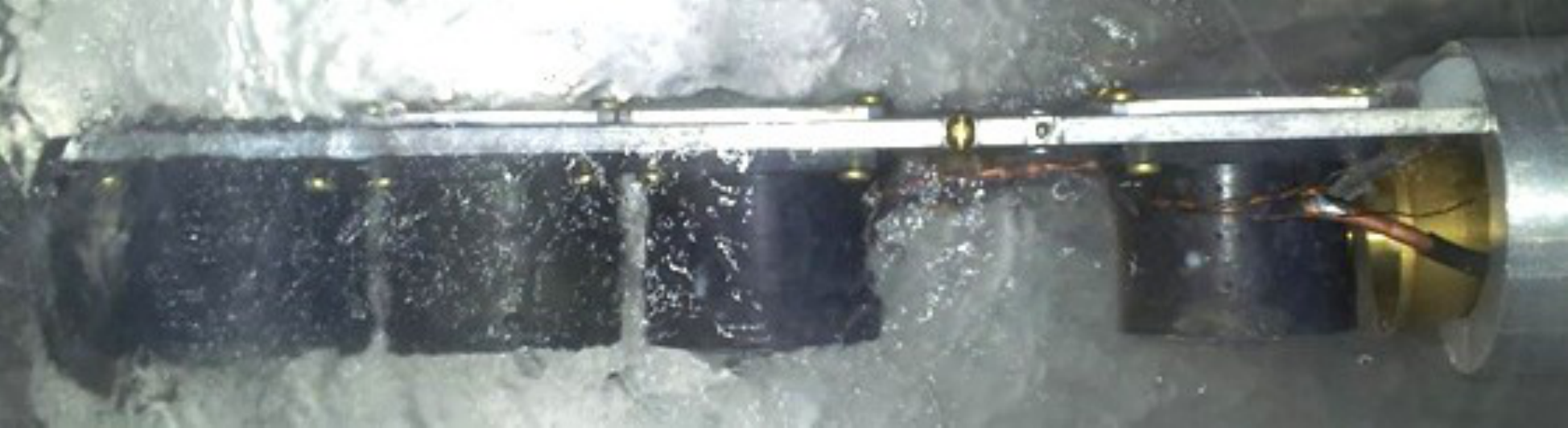}
\end{center}
\caption{The target cells in liquid nitrogen, seen from the side.  The beam would enter from above. The right two cells seen contained the NH$_{3}$ target material.  The microwave horn can be seen at the right of the image.}
\label{ladder}
\end{figure}  

The target insert consisted of a long, 1.65\,in diameter carbon fiber tube with an aluminum ladder piece attached to the end.  This ladder contained the two Kel-f cells for the polarized target material as well as additional target cells for background and beam optics measurements.  The polarized material cells were instrumented with ruthenium oxide resistors for monitoring the temperature during normal operation, and Cernox temperature sensors for monitoring the temperature during cool down and target anneals.  The additional cells could be removed and replaced with cells containing different materials as the need of the experiment changed from energy to energy.  In general, these additional targets consisted of either CH$_{2}$ or carbon targets of various thicknesses, as well as an empty target that contained nothing other than a dummy NMR coil. The ladder portion of the inserts is shown in figure~\ref{ladder}. 

The position of the target cell in the beam's path could be adjusted by raising and lowering the entire insert, which was attached to the top of the refrigerator with an edge-welded bellows.  The lifting is accomplished by means of a stepper motor and linear bearing arrangement, with both the insert and a small platform holding attached equipment moving vertically relative to the refrigerator and therefore the electron beam. The insert has an ISO flange with a stainless steel cylinder welded to the top of it.  The top of the cylinder is sealed with a bolted plate and O-ring.  This plate includes welded, ceramic feedthroughs for the NMR system and the temperature instrumentation, and an O-ring sealed feedthrough for the microwave guide.  The carbon fiber tube that comprises the body of this insert is sufficiently rigid that the alignment of the cell can be accomplished by keying the room temperature vacuum flange.  This allows for a  repeatable and verifiable target orientation, which is important, since the target insert has to be removed on a regular basis to change the target material.

\subsection{NMR} \label{NMR}
\begin{figure}
\begin{center}
\includegraphics[width=2.5in]{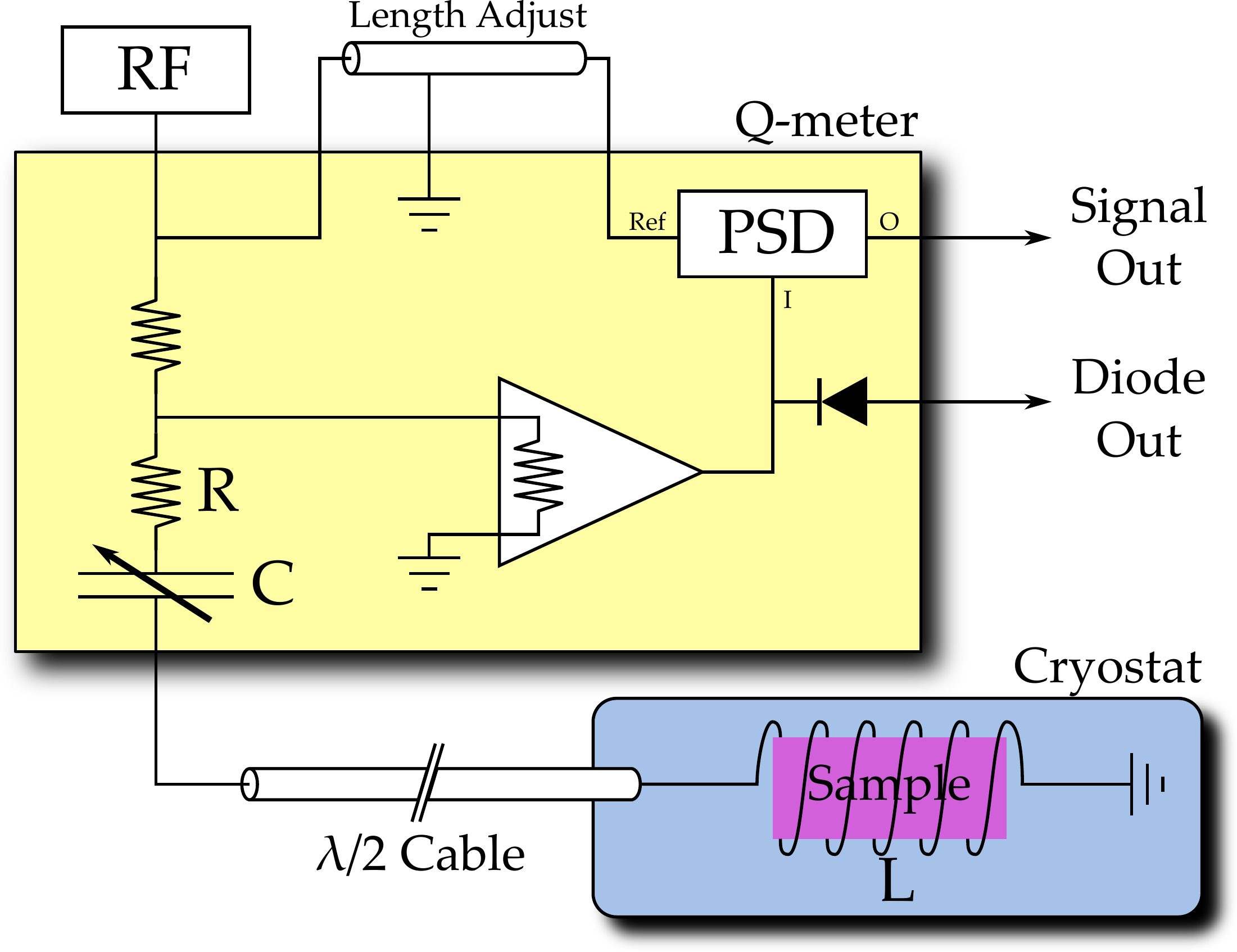}
\end{center}
\caption{A simplified diagram of the NMR circuit}
\label{nmr_fig}
\end{figure}

An NMR system was used to measure the target polarization throughout the experiment.  The NMR system remained basically unchanged from previous DNP targets used at Jefferson Lab and by the University of Virginia Target Group.  The system consists of a series LCR circuit, tuned to the Larmor frequency of the proton, a Liverpool Q-meter~\cite{Court1993}, and a frequency modulating RF power supply.  A diagram of the NMR system is shown in figure~\ref{nmr_fig}.  The inductor of the circuit is a short coil of CuNi capillary installed inside the cell containing the NH$_{3}$ target material.   The output voltage of this system as a function of frequency is digitized using a PC-based ADC.  The area of this function is directly proportional to the polarization of the proton in the target material.  

To calibrate our NMR polarization measurements, 60 thermal equilibrium measurements were taken at 22 occasions throughout the two experiments. Most of these measurements were taken at 1.4 K, although when time allowed additional measurements were made at 1.0 K.

\subsection{Microwaves}
The microwaves necessary for the DNP process where generated by extended interaction oscillator (EIO) tubes\footnote{Communications \& Power Industries LLC}.  Two EIO tubes were used, one for the 2.5\,T target configuration and one for the 5\,T target configuration.    The tubes are matched to the Larmor frequency of the electron, 70\,GHz at 2.5\,T and 140\,GHz at 5\,T.  Each tube has a tunable frequency range of about 1\%.   The EIO tube and associated components were mounted on top of the target lifter platform and moved with the insert.  The 70\,GHz and 140\,GHz tubes and their respective waveguide components were all changed when the magnetic field was changed.  The microwaves were transmitted through a circular waveguide that was attached to the sample insert and were focused on the target material through a horn located above the top target cell.  Only one microwave horn was used, so that the microwaves had to pass through the top cell in order to reach the bottom cell.   About 1\,W of microwave power was delivered to the target cells at at 140\,GHz and more that 2\,W of power at 70\,GHz.  This is due to the lower attenuation in the waveguide at 70\,GHz.

\section{Target performance and results} \label{Results}
The two experiments, $g_2^p$ and $G_E^p$, had different magnetic field and beam current requirements which resulted in varied regimes of target performance. Both experiments took advantage of the higher achievable polarization under a 5\,T magnetic field, although $g_2^p$ took a large portion of its data at 2.5\,T, sacrificing average polarization to achieve the desired acceptance at certain kinematics. Beam heating of the target material generally constrains the beam current to 80--100\,nA, while the sensitivity of the beam position monitoring system put a lower limit on the beam current at roughly 50\,nA during the $g_2^p$ experiment.

The NH$_3$ target material, in the form of 2--3\,mm sized granules, was prepared by irradiation with a 14\,MeV electron beam at the NIST Medical--Industrial Radiation Facility in Gaithersburg, MD. The majority of the material was given an initial dose of approximately $1.2\times10^{17} e/\textrm{cm}^2$ for optimal performance at 2.5\,T, while the remainder was irradiated to approximately $0.9\times10^{17} e/\textrm{cm}^2$ for 5\,T running. The material was irradiated in January of 2012 and stored under liquid nitrogen until use in these experiments.

\subsection{Beam on Target}\label{Anneals}
Roughly $142 \times 10^{15}$ electrons per $\textrm{cm}^2$ (hereafter $\textrm{Pe}/\textrm{cm}^2$) of dose from the CEBAF electron beam were incident on polarized $^{14}$NH$_3$ target samples during these experiments. The proton polarization of pre-irradiated ammonia is observed to decay with the accumulation of additional ionizing radiation at 1\,K \cite{McKee200460}. The detrimental effects of this ``cold dose'' can be temporarily mitigated by warming, or ``annealing'', the target sample.
Using heating elements mounted in the refrigerator, 19 anneals were performed at temperatures between 77 and 97\,K for between 15 and 30 minutes, with the temperature and time increasing with subsequent anneals on a given material.

\begin{figure}
\begin{center}
\includegraphics[width=3.4in]{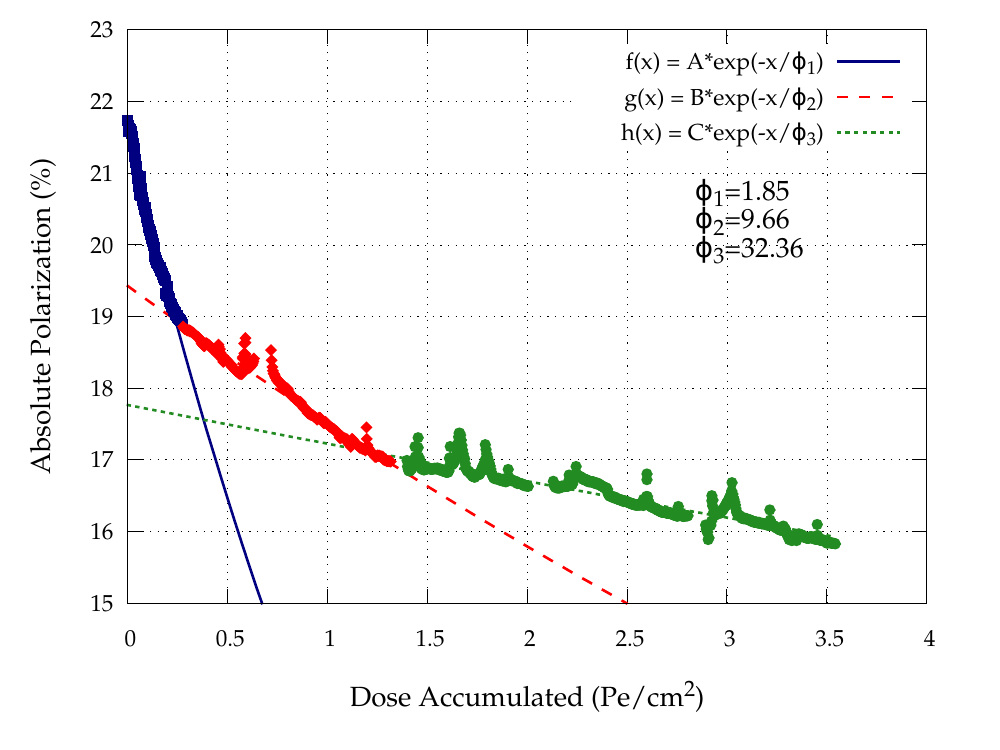}
\end{center}
\caption{Example polarization decay with dose accumulation---from 2.5\,T target field, 80\,nA beam current running---showing three exponential fits to successive decays.}
\label{mat_decay}
\end{figure}

We compare polarization decays vs radiation dose for different field and beam current configurations using exponential fits to the curves: $P(\phi) = P_0 e^{-\phi/\phi_0}$, for radiation dose $\phi$ and so--called ``critical dose'' $\phi_0$. As a decay with dose will typically have more than one exponential, we label the critical dose of each successive exponential as $\phi_1$, $\phi_2$, $\phi_3$, as seen in figure~\ref{mat_decay}.  Table~\ref{dose_tab} shows critical doses for typical decay curves for each configuration, as well as results from Althoff et al. in 1984~\cite{Althoff}.

\begin{table}

\begin{tabular}{l r c c c c}
B Field & Flux ($\textrm{e}/\textrm{cm}^2\textrm{s}$) & $\phi_1$ & $\phi_2$ & $\phi_3$ & P$_\textrm{max}$  \\
\hline
5.0 T &  $15.8 \times 10^{10}$  & 2.4 & 5.5 & 17.2& 92\%\\
5.0 T & $2.0 \times 10^{10}$  & -- & -- &25.6 & 95\%\\
2.5 T & $15.8 \times 10^{10}$  & 1.8 & 9.6 & 32.3 & 28\%\\
2.5 T \cite{Althoff} & $\le 5 \times 10^{10}$ & 1.0 & 4.1 & 30. & 47\%\\

\end{tabular}
\caption{Critical doses for typical polarization decays and maximum achieved polarizations at varied field and beam current settings. Only a single decay constant was seen in the low current, 5\,T setting.}
\label{dose_tab}
\end{table}

\subsection{5 T Field Results}

\begin{figure}
\begin{center}
\includegraphics[width=3.4in]{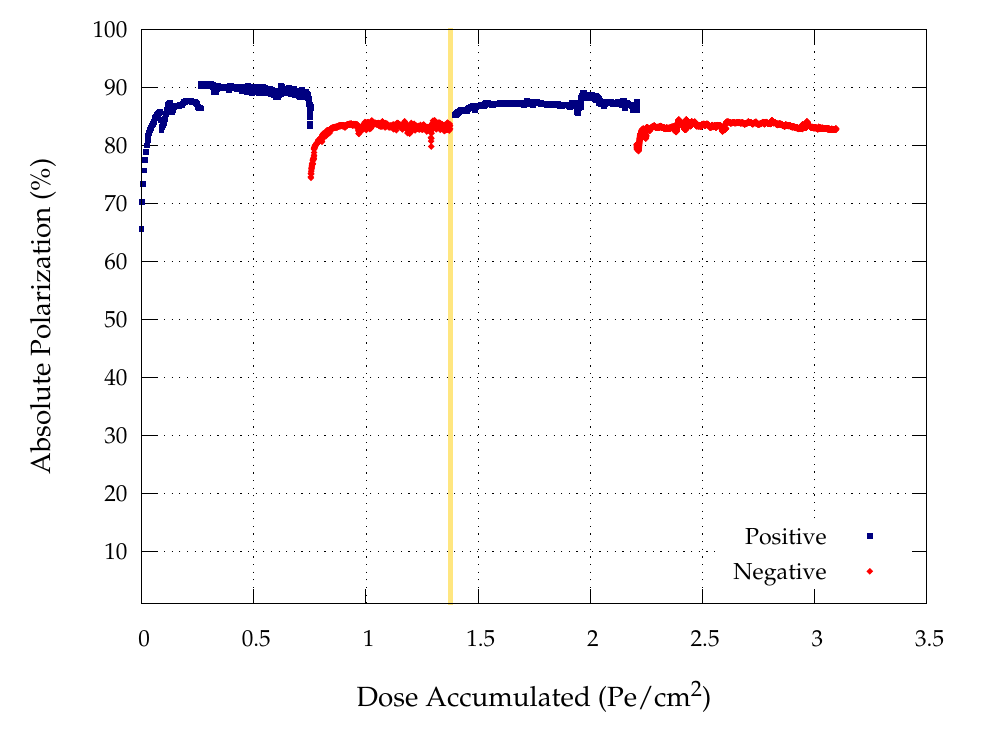}
\end{center}
\caption{Polarization vs. dose for the material which accounted for over half the total dose accumulated during $G_E^p$, taken with a 5\,T magnet field and 10\,nA beam current. The vertical line represents removal and storage at 77\,K.}
\label{gep_mat}
\end{figure}

The polarization performance at the 5\,T target field setting were typical of previous experience; the peak polarization achieved was 95\%.   As is typical in DNP for NH$_{3}$ targets, positive polarization runs outperform the negative by a few percent. When a low electron beam current of 10\,nA was feasible due to the high elastic scattering rates during $G_E^p$, polarizations exceeding 85\% could be maintained for long periods of time, as seen in figure~\ref{gep_mat}. The last decay curve in this figure has a critical dose of 25.6 $\textrm{Pe}/\textrm{cm}^2$. 
The charge--averaged absolute polarization for $G_E^p$, excluding the commissioning irradiation, was 83\%.

At 80 nA beam current and above, the polarization results matched the performance of the 3 previous Hall~C experiments using this target. Anneals were performed after the polarization dropped below roughly 65\%, typically after 2-4\,$\textrm{Pe}/\textrm{cm}^2$ dose on target. An increased decay rate was apparent with subsequent anneals: in a typical example, $\phi_2$ fell from 16.75\,$\textrm{Pe}/\textrm{cm}^2$ after the first anneal to 4.12\,$\textrm{Pe}/\textrm{cm}^2$ after the sixth. Although the peak polarizations often reached 90\%, the charge--averaged polarization for this beam current and field configuration was 70\%.

A strong dependence of the critical dose on the electron beam current was also observed. 
  At the low current, 10\,nA setting, a single decay constant of 25.6\,$\textrm{Pe}/\textrm{cm}^2$ was observed.  This is substantially slower than even the long life decay in the high current case, which was typically around 17.2\,$\textrm{Pe}/\textrm{cm}^2$.   We suspect this indicates additional damage to the polarization due to localized beam heating at higher beam current.

\subsection{2.5 T Field Results}

\begin{figure}[h] 
\begin{center}
\includegraphics[width=3.4in]{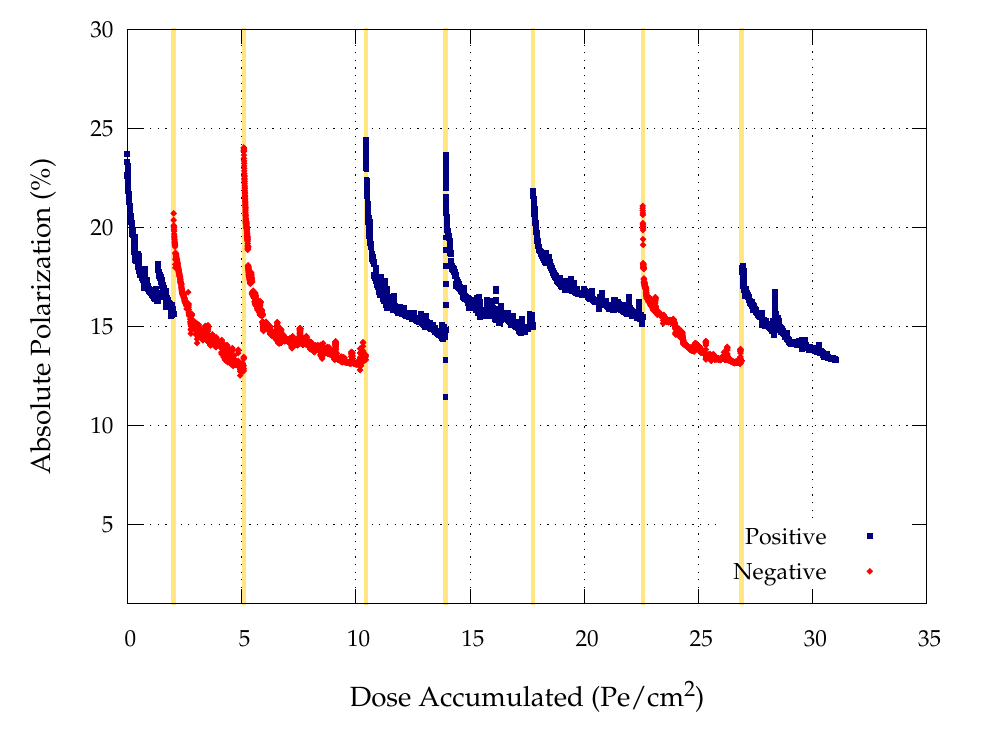}
\end{center}
\caption{Polarization vs. dose for the material which accounted for a third of the total dose accumulated during the $g_2^p$ 2.5\,T magnet field running, under a 80\,nA beam current. Vertical lines represent anneals.}
\label{g2p_mat}
\end{figure}

With DNP under a 2.5\,T magnetic field, much lower peak polarizations are expected; we achieved a maximum 28\% in-beam polarization. 
Figure~\ref{g2p_mat} shows the lifetime of a material sample which accumulated a third of the total dose on target during the 2.5\,T running. The figure shows the 7 anneals on this material, and the decrease in the peak polarization with subsequent anneals.  Although the initial polarization decay after an anneal was stark ($\phi_1 = 1.84 \textrm{Pe}/\textrm{cm}^2$), decay after 17\% was characterized by a very long critical dose ($\phi_3 =32.36 \textrm{Pe}/\textrm{cm}^2$), shown in figure~\ref{mat_decay}. These critcal doses were generally higher than similar results from 1984~\cite{Althoff} seen in table \ref{dose_tab}, although the peak polarizations were lower. The charge--averaged polarization for the 2.5 T running was 15\%.

The T$_1$ relaxation time at 2.5\,T was about 2 minutes, compared to the nearly 30 minutes at 5\,T, reducing the overhead time for thermal equilibrium measurements drastically. 

\section{Summary} \label{Summary}

In order to meet the requirements of both the $g_2^p$ and $G_E^p$ experiments, updates were needed to the DNP target that had been used previously in Hall~C at Jefferson Lab.  The most important updates were a new 1\,K Helium evaporation refrigerator and the addition of a rotating seal to facilitate the rapid rotation of the magnet.  Failure of the Hall~C magnet also required retrofitting the magnet from the Hall~B DNP target to the Hall~C cryostat.  These essential updates, combined with a new target insert, a new alignment system, and a new target positioning system led to very reliable target operation in a variety of configurations.  

Polarizations at 5\,T were on a level consistent with previous operations of DNP targets at Jefferson Lab.  The kinematic requirements of the $g_2^p$ experiment necessitated 2.5\,T running, which resulted in far lower polarizations.  A significant dependence on the beam current was observed in plots of polarization decay as a function of accumulated dose.  This dependence, along with the overhead associated with annealing the target material, changing the sample, and calibrating the NMR should be considered when designing future experiments.  As the figure of merit generally increases as the square of the polarization, in some cases it is possible that a greater overall efficiency could be achieved using lower beam currents.



%% References


\section*{Acknowledgments}
The authors gratefully acknowledge the expert support provided by the technical and engineering 
staffs of Jefferson Lab's Target Group and  Experimental Hall A during the construction, installation, and
operation of this target. We would like to thank Dr.~Fred~Bateman of the MIRF facility at NIST, Gaithersburg for facilitating the irradiation of our ammonia target samples and Dr.~Calvin~Howell, of TUNL at Duke University for the loan of the 70\,GHz EIO tube. We also acknowledge the assistance of Jefferson Lab's Survey and Alignment Group in 
accurately positioning the various target components with respect to
one another and on the beamline.

Authored by Jefferson Science Associates, LLC under 
U.S. DOE Contract No. DE-AC05-06OR23177. 
The U.S. Government retains a non-exclusive, paid-up, irrevocable, world-wide 
license to publish or reproduce this manuscript for U.S. Government purposes.

\bibliographystyle{model1a-num-names}
\bibliography{<your-bib-database>}


\end{document}